\documentclass[12pt]{article}
\usepackage{a4}
\def\R{\mathord{\sf{l\hspace{-0.1em}R}}}

\renewcommand{\title}[1]{\topsep=0pt\begin{center}\Large\bf#1\end{center}\vspace{12pt}}
\renewcommand{\author}[1]{\topsep=0pt\begin{center}\large\rm#1\end{center}}
\newcommand{\institute}[1]{\topsep=0pt\begin{center}\footnotesize\it#1\end{center}\vspace{12pt}}
\renewcommand{\date}[1]{\topsep=0pt\begin{center}\small#1\end{center}\vspace{12pt}}

\begin{document} 

\title{Stability analysis of  cosmological models\\ through Liapunov's
method}

\author{Tiago          C.           Charters          \footnote{email:
charters@cii.fc.ul.pt},    \     Ana    Nunes    \footnote{email:
anunes@lmc.fc.ul.pt},    \   Jos\'e   P.     Mimoso   \footnote{email:
jpmimoso@cii.fc.ul.pt  }}  \institute{$^1$ISEL,  R.   Conselheiro
Em\'{\i}dio  Navarro,1,  1949-014,  Lisboa,  Portugal\\  $^{2,3}$FCUL,
Ed. C1,  Campo Grande,  1749-016 Lisboa, Portugal  \\$^{1,3}$CFNUL and
$^{2}$CMAF, Av. Prof. Gama Pinto, 2, 1649-003, Lisboa, Portugal}

\date{\today}

%%%%%%%%%%%%%%%%%%%%%%%%%%%%%%%%%%%%%%%%%%%%%%%%%%%%%%%%%%%%%%%%%%%%%%

\begin{abstract}
We investigate the general asymptotic behaviour of  Friedman-Robertson-Walker (FRW) models with an inflaton  field,
scalar-tensor  FRW  cosmological  models and  diagonal  Bianchi-IX models  by means of Liapunov's  method. This   method provides information  not  only  about  the  asymptotic stability  of  a  given equilibrium point but also about its basin of attraction.  This cannot be  obtained by the  usual methods  found in  the literature,  such as linear  stability  analysis or  first  order perturbation  techniques. Moreover,  Liapunov's  method  is  also applicable  to  non-autonomous systems.   We use  this  advantadge to  investigate  the mechanism  of reheating for the inflaton field in FRW models.
\end{abstract} 

%%%%%%%%%%%%%%%%%%%%%%%%%%%%%%%%%%%%%%%%%%%%%%%%%%%%%%%%%%%%%%%%%%%%%%

\section{Introduction} 
 
A major difficulty in  analyzing general relativistic problems, namely in  cosmology and black  hole physics,  stems from  the fact  that the relevant  field equations  are  nonlinear. This  seriously limits  the possibility of  obtaining exact solutions,  and makes it  difficult to assess the degree of generality  of the behaviour and special features of those exact solutions which are  known. On the other hand, there is an  increasing  realization  of   the  importance  of  the  asymptotic behaviour  of  models, which  provides  the  relevant  features to  be compared  with  the  physical  data  available for  each  era  of  the universe.
 
As a consequence, in  recent  years the  emphasis  in  the study  of cosmological models has drifted from the search for exact solutions to the analysis of the qualitative properties of the equations and of the long  term behaviour of  the solutions  \cite{Wainright +  Ellis}. The main tools  used in  this context have  been perturbation  methods and qualitative theory, especially  linear stability analysis. Liapunov's method has been  used in    relatively few instances~\cite{Sudarsky:1995,Heusler:1991} as a means to prove stability.
 
In this paper, we argue in  favour of the use of Liapunov's method for the stability analysis of the  possible attractors in a large class of cosmological models,  as an  alternative to linear  approximation. One main advantage of the method  is that it provides information not only about  stability but also  about the  basins of  attraction. Moreover, Liapunov's  method  is   also  applicable  to  non-autonomous  systems~\cite{Rouche+:1977,Hirsch + Smale:1974}.

In its stronger version, Liapunov's method depends on the construction of  a  phase  function  $F$  with the  properties  of  being  strictly decreasing along  the orbits and  having a minimum at  the equilibrium point.   Although  stated  in  most introductory  books  in dynamical systems, this  method is  seldom used in  the applications  because it is difficult in  general to find the  form of the  function $F$. However, there  are some  simple models  in physics  which indicate  an obvious candidate  for $F$  such  as in  the  case of  the classical  textbook application,  the damped  oscillator: the  energy of  the system  is a positive definite form, and decreases along the orbits, thus verifying the condition of applicability of the method in its stronger version.
 
A wide class of cosmological  models share with the latter example the same form and  qualitative dynamics. The equations of  motion for {\it all} these models can be generically written as
\begin{eqnarray} 
\label{ca} 
3\frac{\dot a^2}{a^2} &=& \sum_{i=0}^N \frac{1}{2}\dot 
b_i^2  +  M(a,  b_1,  \ldots,   b_N)  \\  \label{cb}  \ddot  b_i  &=&-
3\frac{\dot a}{a}  \dot b_i -\frac{\partial M}  {\partial b_i}(a, b_1,
\ldots, b_N), \qquad i = 1, 2, \ldots, N
\end{eqnarray} 
where $a$ is non-negative.
 
The  degrees  of freedom  can  be divided  into  two  sets. The  first corresponds to the  global scale factor $a$ which  has, in the simpler models, a monotone behaviour in time,  and the second set is formed by the quantities $b_i$, $i=1,2\ldots, N$,  which in many cases behave as
damped oscillators  and are thus  especially well suited to  the study and analysis through the construction of Liapunov functions.

The paper  is structured as  follows. In Section \ref{Basic  theo}, we briefly  recall  the basic  results  underlying  Liapunov's method. In Section~\ref{Stability analysis} we  shall consider  various examples for  which a  Liapunov  function may  be  constructed that coincides essentially with the  energy of the damped oscillator  modes. For each
case, the relevant dynamical  information provided  by the  method is presented.  First,  in Section~\ref{inflaton}, we study  the basin of attraction of the de Sitter behaviour in Friedman-Robertson-Walker models  with an inflaton field  having a non-vanishing vacuum. We also assess the conditions for the existence of parametric resonance when the scalar field equation exhibits a periodic forcing term. In Section~\ref{reheating} we consider the specific model of  single-field reheating~\cite{Albrecht+:1982,Kofman+:1997,Shtanov+:1995,Kofman+:1994,Kaiser:1996} and show that the results derived on the parametric resonance define the admissible class of potentials.  Next, in  Section~\ref{GR  STGT}, we  consider the  convergence of scalar-tensor gravity  theories to general  relativity in the  case of
Friedman-Robertson-Walker models~\cite{1Damour+Nordtverdt:1993,2Damour+Nordtverdt:1993,Mimoso+Nunes:1998}.
Subsequently, in  Section \ref{Assimp  BIX}, we restrict  ourselves to the simple case of diagonal Bianchi type-IX cosmologies with no matter and a  positive cosmological  constant and study  the validity  of the cosmic no-hair theorem  for this model.  We shall  recover in a simple
manner,  for  this  model,  the  result  proved  in  \cite{Wald:1982}. Finally,  the new  information  provided in  each  case by  Liapunov's method is summed up in Section~\ref{conclu}.

%%%%%%%%%%%%%%%%%%%%%%%%%%%%%%%%%%%%%%%%%%%%%%%%%%%%%%%%%%%%%%%%%%%%%%
\section{Reminder of the basic theorems} 
\label{Basic theo} 
 
Let us start by stating some general definitions.
 
Given a smooth  dynamical system $\dot x = f(x)$,  $x\in \R^n$, and an
equilibrium point $x_0$, we say that a continuous function $F:\R^n \to
\R$, in a neighborhood $U$ of $x_0$, is a Liapunov function for the point $x_0$ if
 
\begin{enumerate} 
 
\item[(i)] $F$ is differentiable in $U\backslash \{x_0\}$,
 
\item[(ii)] $F(x) > F(x_0)$ and
 
\item[(iii)] $\dot F(x)\le 0$ for every $x\in U\backslash \{x_0\}$.
 
\end{enumerate} 
The existence of  a Liapunov function $F$ guarantees  the stability of  $x_0$; furthermore, if the strict  inequality $\dot F(x) < 0$ holds in
$U\backslash  \{x_0\}$ then  $x_0$ is  asymptotically stable  (see for
instance \cite{Hirsch + Smale:1974})

The   neighborhood  $U$  is   essentially  constrained   by  condition
(iii). So, when (iii)  holds strictly, the method provides information
not only  about the asymptotically stability of  the equilibrium point
but also about its basin of  attraction, which must contain the set U.
This  kind  of  information  cannot  be  obtained  neither  by  linear
stability analysis nor by first order perturbation theory.
 
Again  in the  case  when (iii)  holds  strictly, the  existence of  a
Liapunov  function  for  the  equilibrium point  $x_0$  has  important
consequences for the behaviour  of time dependent perturbations of the
above autonomous system. Consider the perturbed system
\begin{equation} 
\dot x = f(x) + g(x,t),
\label{3} 
\end{equation} 
where   $g(x,t)$   is    smooth   and   bounded.    Malkin's   theorem
\cite{Rouche+:1977}  states that the  solutions $x(t;x(t_0),  t_0)$ of
(\ref{3}) will  remain for all  time inside a  prescribed neighborhood
$B_{\epsilon  }$ of $x_0$,  provided that  we pick  initial conditions
$x(t_0)$ sufficiently  close to  $x_0$ and that  the magnitude  of the
perturbation $||g(x,t)||$ is sufficiently small  for all $t > t_0$ and
all $x\in  B_{\epsilon }$.  This  theorem guarantees the  stability of
$x_0$ for  any sufficiently small time dependent  perturbation, but it
does not  imply that  the solution $x(t;x(t_0),  t_0)$ tends  to $x_0$
when  $t \rightarrow \infty  $, since  it does  not even  require that
$g(x,t)$ should  vanish as  $t \rightarrow \infty  $.  If this  is the
case,   then   a   related    result   also   due   to   Malkin   (see
\cite{Rouche+:1977})  shows that an  asymptotic property  persists, in
the sense  that, as time  increases, the neighborhood  $B_{\epsilon }$
may be taken progressively smaller.
 
Liapunov's method can also be used in the more general setting of time
dependent systems  which are  not perturbations of  autonomous systems
(see \cite{Rouche+:1977}). Given  a smooth non-autonomous system $\dot
x =  f(x,t)$, $x\in \R^n$, $t\in  \R$, and an  equilibrium point $x_0$
such that $f(x_0,t)=0$, if  there exists a smooth function $F:U\subset
\R^n  \times  \R \to  \R$,  $U$ a  neighborhood  of  $x_0$, such  that
$F(x_0,t)=F_0=const$,  $F(\cdot ,t)-F_0$ is  positive definite  on $U$
for every $t\in  \R $ and $\dot  F(x,t) \leq 0$ on $U\times  \R $ then
$x_0$ is stable.

This result  is analogous to the  weak version of the  theorem for the
autonomous case.  In the  non-autonomous case, however, the conditions
for  asymptotic stability  are more  restrictive. The  point  $x_0$ is
asymptotically stable  if, moreover,  $F(x,t)$ is bounded  from above,
for  all $t$,  by a  continuous  increasing function  of the  distance
$||x-x_0||$, and $\dot F(x,t)$ is  negative definite in $U$ for all $t
\in \R $.

%%%%%%%%%%%%%%%%%%%%%%%%%%%%%%%%%%%%%%%%%%%%%%%%%%%%%%%%%%%%%%%%%%%%%%
 
\section{Stability analysis} 
\label{Stability analysis} 
 
\subsection{The basin of attraction of de Sitter inflation} 
\label{inflaton} 
 
During  inflation  the leading  contribution  to  the energy  momentum
tensor is given by the  inflaton scalar field $\phi$. The evolution of
the Friedman-Robertson-Walker universes is described by
\begin{eqnarray} 
\label{frwa}
3\frac{{\dot
a}^2}{a^2}+3\frac{k}{a^2}&=&\frac{1}{2}\dot{\phi}^2+V(\phi) \\
\label{frwphi} 
\ddot\phi &=& - 3\frac{\dot{a}}{a}\dot\phi - \frac{{\rm d}V}{{\rm d}\phi}(\phi)
\end{eqnarray} 
where $a$ is the (positive) scale  factor of the universe,   and $k\in\{-1,0,1\}$  distinguishes the
 various spatial curvature cases (notice also that we adopt units in which  $8\pi G=1=c$ throughout the paper). Under  some assumptions on  the scalar field  potential $V$
and on the initial  conditions $\phi_0$, $\dot\phi_0$ and $a_0$, there
exist  solutions where  the  friction term  $3H\dot\phi$ dominates  in
(\ref{frwa})   over   $\ddot  \phi$   and   the   potential  term   in
(\ref{frwphi})      dominates      over      the     kinetic      term
\cite{Kolb+Turner:1989,Charters:2000}.    This  is   the  inflationary
stage,  where the  universe evolves  with accelerated  expansion.  For
definiteness,   we   shall    consider   the   simplest   model   with
$V(\phi)=\lambda + m^2 \phi^2/2$, $\lambda \geq 0$.
 
With  the change  of  variable  $H=\dot a  /a$  and $b=1/a$  equations
(\ref{frwa}) and (\ref{frwphi}) become
\begin{eqnarray} 
\label{Hb1} 
3H^2 &=&-3 k b^{2} + \frac{1}{2}\dot\phi^2 + V(\phi) \\
\label{Hb2} 
\dot b &=& -Hb \\ \ddot\phi &=&- 3H\dot\phi -\frac{{\rm d}V}{{\rm d}\phi}(\phi).
\label{Hb3} 
\end{eqnarray} 
For future use we shall also need
\begin{eqnarray} 
\dot H = k b^2-\frac{1}{2}\dot\phi^2.
\label{use} 
\end{eqnarray} 

Our first goal in this section  will be to apply Liapunov's method for
autonomous systems to the  characterization of the late time behaviour
of  the model  associated  to equations  (\ref{Hb1}), (\ref{Hb2})  and
(\ref{Hb3}). This system  has an equilibrium point at  $(\phi =0, \dot
\phi  =0, b=0)$  corresponding to  $a=\infty $  where $H=\sqrt{\lambda
/3}$.  Since  $H=\dot a/a$, this  equilibrium point corresponds  to an
infinite universe with  exponential asymptotic expansion rate whenever
$\lambda \neq  0$. We shall therefore  call the point  $(\phi =0, \dot
\phi =0, b=0)$ the de Sitter spacetime.
 
Let  us prove  that for  $k=-1,0$ any  initially expanding  model will
approach  the de  Sitter  spacetime asymptotically  for any  $\phi_0$,
$\dot\phi_0$ and $b_0$, and that  for $k=1$ the basin of attraction of
the de Sitter attractor is given by $b_0 < \sqrt{\lambda/3}$.

From (\ref{Hb1}), for an initially  expanding model, $H$ can be written as a function of $b$, $\phi$ and $\dot\phi$ as
\begin{eqnarray} 
H(\phi,\dot\phi                          ,b)                         =
\frac{1}{\sqrt{3}}\left(-3kb^2+\frac{1}{2}\dot{\phi}^2+\lambda        +
\frac{1}{2}m^2\phi^2\right)^{1/2}\; .
\end{eqnarray} 
Notice that the argument of the square root is always positive for $k$
 non-positive. For  the case  $k=1$ this is  also true if  the initial
 value of the scale factor satisfies $b_0 < \sqrt{\lambda/3}$, because
 then $b$ is a monotonically decreasing function of time.
 
The asymptotic stability of the de Sitter attractor is obtained, as in
the  case  of the  classic  damped  oscillator,  by using  a  Liapunov
function given by
\begin{eqnarray} 
\label{lfphi} 
F(b,\phi,\dot\phi) = \frac{3}{2} b^2 + \frac{1}{2}\dot{\phi}^2+V(\phi)
> 0 \; .
\label{liap}
\end{eqnarray} 
From equations (\ref{Hb2}), (\ref{Hb3}) and (\ref{use}) we have
\begin{eqnarray} 
\label{dotflphi} 
\dot F(b,\phi,\dot\phi) = - 3 H \left(\dot\phi^2 + b^2\right) <0,
\end{eqnarray} 
since  $H(\phi,\dot\phi ,  b) >  0$. Then,  $F$ given  by (\ref{liap})
satisfies (i), (ii) and  (iii) for equations (\ref{Hb2}), (\ref{Hb3}),
and (iii)  holds strictly. This  means that for $k=-1,0$  and $\lambda
\geq 0$ the de Sitter spacetime is a global attractor, while for $k=1$
and  $\lambda  >0$  its  basin  of  attraction  contains  all  initial
conditions  $(\phi  _0,   \dot  \phi  _0,  b_0)$  such   that  $b_0  <
\sqrt{\lambda/3}$.
 
Consider now the following modification of the equations of motion
\begin{eqnarray} 
\label{sdfh1} 
3H^2 &=&-3 k b^{2} + \frac{1}{2}\dot\phi^2 + V(\phi) \\
\label{sdfh2} 
\dot b &=& -H b \\
\label{sdfh3} 
\ddot\phi &=&- 3H\dot\phi - m^2\left(1+\varepsilon f(t)\right) \phi,
\end{eqnarray} 
where  $f$ is  a bounded,  continuous, periodic  function  with period
close  to the free  oscillating period  of the  unperturbated equation
(\ref{sdfh3}).

We shall  take equations (\ref{sdfh1}),  (\ref{sdfh2}), (\ref{sdfh3}),
as  a  toy  model for  the  study  of  the  problem of  reheating,  or
parametric  resonance, and  defer to the following subsection its 
application to a specific model.

Because (iii) holds  strictly Malkin's theorem can be  applied to this
case.  It  states that  the solutions of  (\ref{sdfh1}), (\ref{sdfh2})
and  (\ref{sdfh3}),  will remain  for  all  time  inside a  prescribed
neighborhood of  the infinity attractor provided that  we pick initial
conditions  sufficiently close  to it  and that  the magnitude  of the
perturbation  is  sufficiently  small   for  all  $t$.   This  theorem
guarantees   the  stability   of  the   infinity  attractor   for  any
sufficiently small time-dependent  perturbation, but it does not imply
that the solution tends to  it as time goes to infinity. Nevertheless,
we can look  for some hypothesis concerning $H$  which will entail the
asymptotic stability of the equilibrium point $\phi=\dot\phi=b=0$. Let
\begin{eqnarray} 
\label{lftime} 
F(b,\phi,\dot\phi)=    \frac{1}{2}\dot{\phi}^2+\frac{1}{2}\left(3\alpha
H(\phi,\dot\phi    ,b)+m^2\right)\phi^2    +   \alpha\phi\dot\phi    +
\frac{3}{2}b^2,
\end{eqnarray} 
which  is  positive definite  for  $0<\alpha  <m$.   We obtain,  using
(\ref{sdfh1}), (\ref{sdfh2}), (\ref{sdfh3}),
\begin{eqnarray} 
\label{td} 
\dot       F        =       -\left(3H       -\alpha\right)\dot{\phi}^2
&-&\alpha\left(m^2-\frac{3}{2}\dot       H       +m^2      \varepsilon
f(t)\right)\phi^2  -  m^2\varepsilon  f(t)\phi\dot\phi\nonumber  \\&-&
3Hb^2.
\end{eqnarray} 

$\dot F$ will be negative definite if
\begin{eqnarray} 
\label{inequal}
\left(3H  -\alpha\right)\left(m^2-\frac{3}{2}\dot  H +  m^2\varepsilon
f(t)\right)-\frac{1}{4}m^4\varepsilon^2 f^2(t)>0.
\end{eqnarray} 
%This condition is satisfied for any small enough $\varepsilon$. 
Since $\dot H(\phi =0, \dot \phi =0, b=0)=0$, this condition will hold
for  $t$   large  enough  and   $\epsilon  $  small   enough  whenever
$\displaystyle \lim _{t\rightarrow +\infty } H(t) >\alpha /3$.

There  are two  basic  conclusions  that can  be  obtained from  these facts. First, parametric resonance can only occur when (\ref{inequal}) does not  hold, that  is, parametric resonance  can only occur  if $H$
tends  to zero  as  time  goes to  infinity,  because, otherwise,  the Liapunov function  (\ref{lftime}), ensures that  the equilibrium point $\phi=\dot\phi=b_0=0$  is asymptotically  stable.   This implies,  for this toy model,  that reheating can only be  obtained from a potential that has a null minimum $V(0)=\lambda = 0$.  Second, even if $H$ tends to zero  as time  goes to infinity,  parametric resonance can  only be observed if $\varepsilon$ is  greater than some critical value because
otherwise  Malkin's  theorem implies  the  stability  of the  infinity attractor.

%%%%%%%%%%%%%%%%%%%%%%%%%%%%%%%%%%%%%%%%%%%%%%%%
\subsection{Single field reheating}
\label{reheating}

We  now  examine a specific model of reheating after inflation. According  to  the inflationary theory,  almost  all elementary  particles populating the  universe were created during the  process of reheating after inflation~\cite{Kofman+:1997}.  Reheating occurs due to particle  production  by the oscillating  scalar field  $\phi$. In  the simplest
models,  this field  $\phi$ is  the  same inflaton  field that  drives inflation at the early stages of the evolution of the universe.  After inflation, the  inflaton oscillates near the minimum  of its potential and this induces the quantum process of creation of particles \cite{Kofman+:1997}. At the root of this process is a phenomenon of parametric resonance which gives rise to the excitation of certain modes of the quantum fluctuations of fields \cite{Kofman+:1994,Traschen+:1990}. 

We shall consider a simple model  with the scalar field potential of the form $V(\phi)=V_0\phi^n$, with $V_0>0$, characteristic of the chaotic inflation scenario, and of the original studies of reheating. Our purpose is to ilustrate the possibilities underlying the aplication of the results of the previous subsection to this model. We aim at determining under what  conditions does the mechanism of resonance occur, avoiding the drastic assumption often made in the analytic treatment found in the literature which consists in neglecting  the Hubble rate of expansion~\cite{Albrecht+:1982,Kofman+:1997,Kofman+:1994,Kaiser:1996}. Furthermore, we also wish to show that the theorem on parametric resonance requiring the damping term to be vanishing in time and the amplitude of the periodic perturbation in the scalar field equation to be bounded from below permits to constrain the form of the potential in the case under consideration.

We  consider a FRW  metric with non-positive curvature and
shall focus on perturbations of the single scalar field ignoring the accompanying metric fluctuations. This is well justified on scales smaller than the Hubble distance, that is, satisfying $p\ge a\,H$, where $p$ is the wave number of the perturbation, since relativistic effects play a lesser role on these scales \cite{Mukhanov:1992+}. On scales larger than the Hubble radius this approximation is more restrictive and, in particular, it has been recently shown to miss the resonant behaviour of the metric perturbations \cite{Bassett+:1999a,Bassett+:1999b,Finelli+:1999}. With this caveat we shall thus  consider the set of equations of motion 
\begin{eqnarray}
3H^2 &=&-3 k b^{2} + \frac{1}{2}\dot\phi^2  + V(\phi) \\ \dot b &=& -H
b       \\       \ddot\phi       &=&-       3H\dot\phi       -       n
V_0\phi^{n-1}+b^2\nabla_{\vec x}^2\phi,
\end{eqnarray} 
with $n > 1$ and $\vec x $ the spatial coordinates. 

Following~\cite{Kaiser:1996}, we decompose the inflaton field into the
sum of a classical inflaton  field $\varphi$ and a quantum fluctuation
$\delta\phi$ in the form $\phi (\vec  x,t) = \varphi (t) + \delta \phi
(\vec x,t)$. We next concentrate in a specific mode of the fluctuation
$\delta \phi  $. This yields,  keeping the first
order terms in $\delta\phi_p$,
\begin{eqnarray}
&&\ddot\varphi   +  3H\dot\varphi   +  n   V_0\varphi^{n-1}  =   0  \\
&&\delta\ddot\phi_p    +    3H\delta\dot\phi_p    +   \left[n    (n-1)
V_0\varphi^{n-2}+p^2 b^2\right]\delta\phi_p = 0.
\end{eqnarray} 
Notice that $n$ should be  strictly greater than two for the existence
of resonant behaviour.

Let $d\tau/dt = b$ and denote the derivative $d/d\tau$ by a prime. The
equations of motion become
\begin{eqnarray}
3H^2 &=& -3kb^2+\frac{1}{2}b^2\varphi'^2 + V(\varphi) \\ b' &=& -H\\
\label{qper1} 
\varphi'' &=&2\frac{b'}{b}\varphi' - n b^{-2} V_0\varphi^{n-1} \\
\label{qper}
\delta \phi_p'' &=&  2\frac{b'}{b}\delta\phi_p' - \left[n (n-1) b^{-2}
V_0\varphi^{n-2}+p^2\right]\delta\phi_p
\end{eqnarray} 
with $n>2$.

Let   $T=\dot\varphi^2/2=b^2\varphi   '^2/2   $   and  $\gamma   =   2
T/(T+V)$. Since  $2\overline T =  n \overline V$  holds asymptotically
\cite  {Turner:1983}, we  have  $\overline \gamma  =  2n/(n+2)$ and  a
simple argument  shows that $H\sim b^{3\overline  \gamma/2}$ if $k=0$,
$H\sim  b$ if $k=-1$.   Consequently $b'/b$  and $b^{-2}\varphi^{n-2}$
approach the infinity attractor as
\begin{eqnarray}
b'/b &\sim& \cases{b^{(2n-2)/(n+2)} &if $k=0$ \cr b &if $k=-1$ }
\label{fin1}\\ \varphi^{n-2}b^{-2} &\sim& b^{(4n-16)/(n+2)}. 
\label{fin2}
\end{eqnarray} 

From the previous  discussion of our toy model we  know that, in order
to have  parametric resonance, the  damping term in  (\ref{qper}) must
tend to zero  and the perturbing term in  (\ref{qper}) must be bounded
away from  zero. The first condition  is always satisfied,  in view of
(\ref{fin1}) and of the fact that  $n$ must be greater than $2$.  From
(\ref{fin2}), the  second condition is satisfied  provided that $n\leq
4$.   Notice   that  for  $n=4$  the   term  $\varphi^{n-2}b^{-2}$  in
(\ref{qper}) oscillates with constant  amplitude for all time. This is
the  case   usually  studied  in  the  literature   (see  for  example
\cite{Kofman+:1997}).   However,  (\ref{fin2})  shows  that  reheating
should be more effective for  potentials with exponents closer to 2 in
the allowed interval $(2,4]$.

%%%%%%%%%%%%%%%%%%%%%%%%%%%%%%%%%%%%%%%%%%%%%%%%%%%%%%%%%%%%%%%%%%%%%%

\subsection{General Relativity as a Cosmological attractor \\ 
of Scalar-Tensor Gravity Theories}
\label{GR STGT} 

General scalar-tensor gravity theories are based on the Lagrangian
\cite{1Damour+Nordtverdt:1993,2Damour+Nordtverdt:1993,Mimoso+Nunes:1998}
\begin{eqnarray} 
L_\Phi                    =                   \Phi                   R
-\frac{\omega(\Phi)}{\Phi}g^{\mu\nu}\Phi_{,\mu}\Phi_{,\nu}  + 2U(\Phi)
+ 16\pi L_m \label{ST_action}
\end{eqnarray} 
where $R$ is the Ricci curvature scalar of a space time endowed with a
metric  $g_{\mu\nu}$, $\Phi$ is  a scalar  field, $\omega(\Phi)$  is a
dimensionless coupling  function, $U(\Phi)$  is a function  of $\Phi$,
and $L_m$ is the Lagrangian for the matter fields. They provide a most
natural  generalization of  Einstein's general  relativity,  and their
investigation enables a generic model-independent approach to the main
features  and cosmological implications  of the  unifications schemes.
The   archetypal    of   these   theories    is   Brans-Dicke   theory
\cite{Brans+Dicke:1961,Dicke:1962}  for   which  $\omega(\Phi)$  is  a
constant and the  most distinctive feature of the  theories defined by
(\ref{ST_action})  is  the variation  of  the gravitational  constant,
since $G=\Phi^{-1}$.
 
Consider the homogeneous and isotropic Friedman-Robertson-Walker (FRW)
universes  and assume that  the matter  content of  the universe  is a
perfect fluid with the usual  equation of state $p =(\gamma -1 )\rho$,
where $0\le\gamma\le  2$. Here $p$  and $\rho$ are,  respectively, the
pressure  and  the energy-density  measured  by  a comoving  observer.
Performing    the    conformal    transformation   of    the    metric
\cite{Dicke:1962,Mimoso+Wands:1995}
$\tilde{g}_{ab}=(\Phi/\Phi_*)\,g_{ab}$,  where $\Phi_*$ is  a constant
which normalizes the gravitational constant, to the so-called Einstein
frame    which     corresponds    to  replacing   the dynamical variables $(a,t,\Phi)$ with
$\left(\sqrt{\frac{\Phi}{\Phi_*}}   a,  \sqrt{\frac{\Phi}{\Phi_*}}  t,
\varphi\right)$ where
\begin{eqnarray} 
\frac{d\ln\Phi}{d\varphi}=\sqrt{\frac{16\pi}{\Phi_*(2\omega(\varphi)+3)}}
\label{alpha} \; ,
\end{eqnarray} 
the field equations reduce to the simple form
\begin{eqnarray} 
3\frac{{\dot
a}^2}{a^2}+3\frac{k}{a^2}&=&\frac{1}{2}\dot{\varphi}^2+a^{-3\gamma}M(\varphi)+V(\varphi)
\\
\label{eqnscf} 
\ddot\varphi              +              3\frac{\dot{a}}{a}\dot\varphi
&=&-a^{-3\gamma}M'(\varphi)-\frac{{\rm d}V}{{\rm d}\varphi}(\varphi),
\end{eqnarray} 
where
$V(\varphi)=\Phi^2_*U\left(\Phi(\varphi)\right)/\left(8\pi\Phi^2(\varphi)\right)$,
$M(\varphi)=\mu\left(\Phi(\varphi)/\Phi_*\right)^{-2+3\gamma/2}$    and
$\mu$  is  a constant  which  fixes  the  initial conditions  for  the
energy-density of the perfect  fluid. We shall assume that $M(\varphi)
\ge   0$   (which  corresponds   to   having   $G\ge   0$)  and   that
$V(\varphi)=\lambda+\frac{1}{2}m^2\varphi^2$,  i.e., that $V(\varphi)$
has a non-degenerate positive global minimum at the origin.
 
Let $H=\dot  a / a$ and  $b=1/a$.  It is  easy to see that  the former
system is equivalent to
\begin{eqnarray} 
\label{Heq} 
3H^2 &=&-3 k  b^{2} + \frac{1}{2}\dot\varphi^2 + b^{3\gamma}M(\varphi)
+ V(\varphi) \\
\label{beq} 
\dot b &=& -Hb \\
\label{varphieq} 
\ddot\varphi &=&- 3H\dot\varphi - b^{3\gamma}M'(\varphi)-\frac{{\rm d}V}{{\rm d}\varphi}(\varphi).
\end{eqnarray} 
 
These equations  define a dynamical  system of dimension $3$.   We are
 interested in studying the stability of the equilibrium point at $a=+
 \infty   $   corresponding   to   $b=0$,   $H=\sqrt{\lambda/3}$   and
 $\varphi=\dot\varphi  =   0$,  which   is  the  de   Sitter  solution
 \cite{Mimoso+Nunes:1998}.
 
Let  us prove  that for  $k=-1,0$ any  initially expanding  model will
approach the  de Sitter spacetime asymptotically  for any $\varphi_0$,
$\dot\varphi_0$ and $b_0$, and that  for $k=1$ the basin of attraction
of  the de  Sitter attractor  is  given by  $b_0 <  \sqrt{\lambda/3}$.
Consider the  initial conditions $H_0>0$ associated  with an expanding
model,  and $b_0$, $\varphi_0$  and $\dot\varphi_0$  arbitrary.  Using
the   hypothesis   $M(\varphi   )\geq   0$,   the   expanding   branch
$H=H(\varphi,\dot\varphi ,b)$ of (\ref{beq})is given by
\begin{eqnarray} 
\label{Heqsqrt} 
H =\frac{1}{\sqrt{3}}  \left\{-3 k b^{2}  + \frac{1}{2}\dot\varphi^2 +
b^{3\gamma}M(\varphi) + V(\varphi)\right\}^{1/2}>0 \; .
\end{eqnarray} 
 
As  for   the  case   $k=1$,  we  take   again  $H_0>0$   and  $b_0\le
\sqrt{\lambda/3}$  and, since,  from  (\ref{beq}), $b$  is a  strictly
decreasing  function  of  time,  it  is  possible  for  these  initial
conditions    to   write    $H=H(\varphi,\dot\varphi   ,b)$    as   in
(\ref{Heqsqrt}).
 
The function defined by
\begin{eqnarray} 
\label{fliapunov} 
F(b,\varphi,\dot\varphi)    =    \frac{1}{2}\dot\varphi^2+\lambda    +
\frac{1}{2}m^2\varphi^2 + b^{3\gamma}M(\varphi)
\end{eqnarray} 
is    a    Liapunov     function    for    the    equilibrium    point
$(b,\varphi,\dot\varphi)=(0,0,0)$.     Indeed    using    (\ref{Heq}),
(\ref{beq}), (\ref{varphieq})  and (\ref{Heqsqrt}), it is  easy to see
that
\begin{eqnarray} 
\dot   F(b,\varphi,\dot\varphi)  =  -3H\left(\dot\varphi^2   +  \gamma
b^{3\gamma}M(\varphi)\right) < 0 \; ,
\end{eqnarray} 
and it  is also straightforward  to check that  (i) and (ii)  hold for
$F$.     Thus    this    proves    that    the    equilibrium    point
$(b,\varphi,\dot\varphi)=(0,0,0)$  is asymptotically stable  and shows
that an attractor  mechanism indeed exists when the  matter content of
the universe is a perfect fluid and the potential has a non-degenerate
non-negative minimum.  This means  that for $k=-1,0$ and $\lambda \geq
0$ the de Sitter spacetime is  a global attractor, while for $k=1$ and
$\lambda >0$  its basin of attraction contains  all initial conditions
$(\varphi   _0,   \dot   \varphi   _0,   b_0)$  such   that   $b_0   <
\sqrt{\lambda/3}$. A similar attracting mechanism should also exist in
the non-isotropic case.
 
In general, linear stability  analysis only provides information about
the stability and asymptotic  behaviour in a undetermined neighborhood
of the  equilibrium point.  This can  be easily seen  in the following
example. Consider   the    equations   of    motion   (\ref{beq}),
(\ref{varphieq})  for $\gamma=4/3$  (for $\gamma\neq  4/3$  the linear
stability method  can not be applied without  additional hypothesis on
$M(\varphi)$). The eigenvalues associated with the de Sitter attractor
are    $\lambda_b    =    -\sqrt{\lambda/3}$,    $\lambda_\varphi    =
-\sqrt{3\lambda}-\sqrt{3\lambda  -4m^2}$ and  $\lambda_{\dot\varphi} =
-\sqrt{3\lambda  }+\sqrt{3\lambda -4m^2}$.   The  linear approximation
confirms the result obtained by  the Liapunov method, the real part of
the eigenvalues  is always  negative, and shows  that there  exist two
basic behaviours for the field  associated to the monotone approach of
the  scale factor  to the  de  Sitter attractor:  an oscillatory  mode
(damped oscillations) if $0<\lambda < 4m^2/3$ and a monotone behaviour
(over-damped) for the field if $\lambda \ge 4m^2/3$ which corresponds,
respectively,  to  a stable  focus  and a  stable  node  in the  plane
$(\varphi, \dot\varphi,  b=0)$. No information  is obtained concerning
the size of  the basin of attraction for the  de Sitter attractor with
this method.
 
Also  notice that  when $\gamma  <1/3$ the  second member  of equation
(\ref{varphieq}) is  not differentiable at $b=0$,  which precludes the
application of  linear stability analysis  in its simplest  form. This
fact emphasizes the advantage of the use of the Liapunov's method.

%%%%%%%%%%%%%%%%%%%%%%%%%%%%%%%%%%%%%%%%%%%%%%%%%%%%%%%%%%%%%%%%%%%%%% 

\subsection{Asymptotic behaviour for diagonal Bianchi type-IX 
cosmologies  in  the  presence  of a  positive  cosmological  constant
without matter.}
\label{Assimp BIX} 
 
 By definition, the general Bianchi type-IX spacetime has topology $\R
 \times {\bf{S}}^3$,  with a simply transitive action  of the isometry
 group  $SU(2)$  on ${\bf{S}}^3$  spatial  slices.   The  metric of  a
 general Bianchi  type-IX model  can be put  in the form  \cite{Ryan +
 Shepley:1975}
\begin{eqnarray} 
ds^2          =          -dt^2         +          a^2(t)\sum_{i,j=1}^3
\left[{e^{\beta}}\right]_{ij}d\omega^i d\omega^j \; ,
\end{eqnarray} 
where $\omega^i$,  $i=1,2,3$, are isometry invariant  one-forms on the
three-sphere, $a$ is a scalar,  and $\beta$ is a traceless $3\times 3$
matrix.  Both  $a$ and  $\beta$ are functions  of the proper  time $t$
only.
 
For  a  diagonal  spacetime,  let  $\beta_i$,  $i=1,2,3$,  denote  the
diagonal elements of the matrix  $\beta$. Only two of these quantities
are independent since $\beta_1 + \beta_2  + \beta_3 = 0$ on account of
the tracelessness  of $\beta$. We choose the  independent variables to
be \cite{Lin:1989}
\begin{eqnarray} 
b_1       &=&-\frac{1}{2\sqrt{6}}\beta_3        \\       b_2       &=&
\frac{1}{6\sqrt{2}}\left(\beta_1-\beta_2\right).
\end{eqnarray} 
 
The vacuum Einstein  equations with the cosmological constant take  the form
\cite {Ryan + Shepley:1975,Lin:1989}
\begin{eqnarray} 
\label{a} 
3\frac{\dot a^2}{a^2}  = \frac{1}{2}\left(\dot b_1^2+\dot b_2^2\right)
+ \frac{1}{a^2}V(b_1,b_2) + \Lambda
\end{eqnarray} 
\begin{eqnarray} 
\label{b1b2} 
\ddot  b_i + 3\frac{\dot  a}{a}\dot b_i  + \frac{1}{a^2}\frac{\partial
V}{\partial b_i} = 0, \qquad i = 1,2
\end{eqnarray} 
with
\begin{eqnarray} 
\label{Vb1b2} 
V(b_1,b_2)      &=&      -e^{-\sqrt{\frac{2}{3}}b_1}\cosh(\sqrt{2}b_2)
+\frac{1}{4}e^{-4\sqrt{\frac{2}{3}}b_1} \\  \nonumber & &\qquad \qquad
+
\frac{1}{2}e^{2\sqrt{\frac{2}{3}}b_1}\left[\cosh(2\sqrt{2}b_2)-1\right]
\end{eqnarray} 
 
Let $H=\dot  a / a$ and  $b=1/a$.  It is  easy to see that  the former
system is equivalent to
\begin{eqnarray} 
3H^2  &=&\frac{1}{2}\left(\dot   b_1^2+\dot  b_2^2\right)  +\Lambda  +
b^2V(b_1,b_2) = 0 \\
\label{bbeq} 
\dot b &=& -H b
\end{eqnarray} 
\begin{eqnarray} 
%\label{b1b2} 
\ddot  b_i +  3H\dot b_i  +  b^2\frac{\partial V}{\partial  b_i} =  0,
\qquad i = 1,2.
\end{eqnarray} 
These  equations define  a dynamical  system of  dimension  $5$.  This
system has  two finite equilibrium  points $b=0$, $H=\sqrt{\Lambda/3}$
and $b=2\sqrt{\Lambda/3}$, $H=0$ both with $b_1=b_2=\dot b_1= \dot b_2
=0$. We are interested in the stability of the equilibrium point $b=0$
associated with the de Sitter attractor.
 
Notice   that  near  the   minimum  $(b_1,b_2)=(0,0)$   the  potential
(\ref{Vb1b2}) takes the form
\begin{eqnarray} 
V(b_1,b_2) &=& -\frac{3}{4}  + \left(b_1^2+b_2^2\right) + O_3(b_1,b_2)
	   \nonumber \\ &=& -\frac{3}{4} + v(b_1,b_2),
\end{eqnarray} 
where $v(0,0)=0$.
 
We   take  $H_0>0$  and   $b_0\le  2\sqrt{\Lambda/3}$.    Since,  from
(\ref{bbeq}), $b$ is a strictly monotone decreasing function with time
the  expanding branch  of  $H=H(b,b_1,b_2,\dot b_1,\dot  b_2)$ can  be
written for these initial conditions as
\begin{eqnarray} 
\label{HeqsqrtBIX} 
H   =\frac{1}{\sqrt{3}}   \left\{   \frac{1}{2}\left(\dot   b_1^2+\dot
b_2^2\right) +  b^2v(b_1,b_2)+\Lambda -\frac{3}{4}b^2 \right\}^{1/2}>0
\; .
\end{eqnarray} 
 
The  Liapunov  function  for the equilibrium  point  $(b,b_1,b_2,\dot
b_1,\dot b_2)=(0,0,0,0,0)$ will now be
\begin{eqnarray} 
\label{fliapunovBIX} 
F(b,b_1,b_2,\dot  b_1,\dot  b_2)  =  \frac{1}{2}\left(\dot  b_1^2+\dot
b_2^2\right)+ b^2v(b_1,b_2) + \frac{3}{2} b^2 \ge 0 \; .
\end{eqnarray} 
Indeed using the equations of motion we have
\begin{eqnarray} 
\dot      F      =      -3H\left[\dot     b_1^2+\dot      b_2^2      +
\frac{2}{3}b^2v(b_1,b_2)+b^2\right] < 0
\end{eqnarray} 
and thus it  is a simple matter to check that  all the conditions (i),
(ii)  and  (iii)  are  satisfied.   Therefore  the  equilibrium  point
$(b,b_1,b_2,\dot  b_1,\dot b_2)=(0,0,0,0,0)$ is  asymptotically stable
and  the basin  of attraction  for  de Sitter  spacetime contains  all
initial     conditions    $\left(b_0,b_1(0),b_2(0),\dot    b_1(0),\dot
b_2(0)\right)$ such that $b_0  <2 \sqrt{\Lambda/3}$.  In this case the
linear stability analysis reveals only  that the eigenvalues of the de
Sitter attractor are all negative.

%%%%%%%%%%%%%%%%%%%%%%%%%%%%%%%%%%%%%%%%%%%%%%%%%%%%%%%%%%%%%%%%%%%%%%

\section{Summary and conclusions} 
\label{conclu} 
 
In this  paper we have applied  the Liapunov's method to  the study of
the general  asymptotic behaviour of cosmological  models, namely: FRW
cosmologies  with inflaton  field; scalar-tensor  gravity  theories in
FRW;  diagonal Bianchi-IX  vacuum  cosmologies in  the  presence of  a
positive cosmological  constant.  For these models  we recover through
this method the well known stability result for the de Sitter solution
and we show that the  method also provides information about its basin
of attraction.

 Furthermore, we have applied  Liapunov's method to another problem of
a completely  different nature, namely the analysis  of the conditions
for the existence of the parametric resonance mechanism underlying the
reheating process.  We have concluded that in order to have parametric
resonance, the damping term  in the quantum perturbation equation must
tend to zero and the forcing term must be bounded away from zero. This
implies  that  for  the  basic  model  with  potentials  of  the  form
$V(\varphi)=V_0\varphi^n$, reheating may occur only when $n\in (2,4]$.
As far as we know, albeit the simplicity of the reheating model considered, this conclusion is new and fully illustrates the extension to which Liapunov's method may be utilized to draw rigorous results in cosmology.

%%%%%%%%%%%%%%%%%%%%%%%%%%%%%%%%%%%%%%%%%%%%%%%%%%%%%%%%%%%%%%%%%%%%%%

\section*{Acknowledgements}
The  authors are grateful to Alfredo B. Henriques for useful discussions and wish to  acknowledge the  finantial support  from Funda\c c\~ao de Ci\^encia e Tecnologia under the grant  PBIC/C/FIS/2215/95.

%%%%%%%%%%%%%%%%%%%%%%%%%%%%%%%%%%%%%%%%%%%%%%%%%%%%%%%%%%%%%%%%%%%%%%

\end{document}